# Application of Deep Learning in Recognizing Bates Numbers and Confidentiality Stamping from Images

Christian J. Mahoney
e-Discovery
Cleary Gottlieb Steen & Hamilton LLP
Washington, D.C. USA
Email: cmahoney@cgsh.com

Katie Jensen
Data & Technology
Ankura Consulting Group, LLC
New York, NY USA
Email: katie.jensen@ankura.com

Fusheng Wei
Data & Technology
Ankura Consulting Group, LLC
Washington, D.C. USA
Email: fusheng.wei@ankura.com

Haozhen Zhao
Data & Technology
Ankura Consulting Group, LLC
Washington, D.C. USA
Email: haozhen.zhao@ankura.com

Han Qin
Data & Technology
Ankura Consulting Group, LLC
Washington, D.C. USA
Email: han.qin@ankura.com

Shi Ye
Data & Technology
Ankura Consulting Group, LLC
Washington, D.C. USA
Email: shi.ye@ankura.com

*Abstract*— **In eDiscovery, it is critical to ensure that each page produced in legal proceedings conforms with the requirements of court or government agency production requests. Errors in productions could have severe consequences in a case, putting a party in an adverse position. The volume of pages produced continues to increase, and tremendous time and effort has been taken to ensure quality control of document productions. This has historically been a manual and laborious process. This paper demonstrates a novel automated production quality control application which leverages deep learning-based image recognition technology to extract Bates Number and Confidentiality Stamping from legal case production images and validate their correctness. Effectiveness of the method is verified with an experiment using a real-world production data.**

## I. INTRODUCTION

In 2018, the Court Statistics Project reported 17 million incoming Criminal cases and 16.4 million Civil in State Courts were opened [1]. If every case one of those cases included a discovery component, that's a potential for 33.4 million document requests. A legal document request is a formal notice used during the Electronic Discovery (eDiscovery) phase from another party asking for copies of documents or files that are relevant to litigation or an investigation. A document request includes guidance on both what data to identify and collect as well as how to respond to the request by providing copies of relevant data. Norton Rose Fullbright's 2019 Litigation Survey suggests that 35% of respondents "expect volume of disputes to rise moving forward"[2]. More disputes correlate to more document requests.

To comply with a document request, a company identifies and collects electronically stored information (ESI) such as emails, word documents, and excel spreadsheets. That data is put through a review process to identify the set of relevant material. Data is segregated into a set of documents that will be turned over and those that will not. The data that is identified as relevant for the case then moves through a production process, based upon a set of rules and requirements as set out in the document request.

The widely recognized standard for document productions, as described by the Securities and Exchange Commission, (SEC), is that "Documents must be uniquely and sequentially Bates-numbered across the entire production, with an endorsement burned into each image."[3]. Each document is printed to one or more image files, depending on the number of pages of the original, capturing the full contents of the file. The term Bates Stamp comes from several patents acquired by Edwin G. Bates which were filed for a machine that was able to apply consecutive numbers to paper files, where the stamping machine would rotate one letter or number after each application.

Today's use of Bates number is moving further and further away from applying numbers to paper documents. Instead, electronic files are turned into "near-paper" files, or static images so that altercations cannot be made. The format of the image is often in a Portable Document Format (PDF) or Tag Image File Format (TIFF). The Bates number is typically a combination of letters, as a prefix, followed by numbers. The Bates number for each page is added to that static image on the lower right-hand side and each page marked sequentially. The combination of a static image with a unique identifier makes it easy for Legal Counsel to reference an exact page within a larger production set and used in responses to inquiries, depositions and/or court proceedings.

A Bates number is not the only thing that is applied on images or pages within a document production. The lower left-hand corner of a produced image can be used to convey different Confidentiality treatments in compliance with Protective Orders typically negotiated on Large Litigation cases and government investigations. One purpose of a Protective Order is to govern how produced document can be utilized and who can be given access to particular produced documents.

Government agencies rely on Confidentiality stamping of pages and documents when responding to requests under the Freedom of Information Act, (FOIA) (5 U.S.C. § 552). FOIA, enacted in 1996, "has provided an important means through which the public can obtain information regarding the activities of Federal agencies"[4]. The SEC received 11,546 FOIA requests in 2019 [5].

The government is required to provide copies of all documents produced to an agency unless it qualifies for an exemption under (5 U.S.C. § 552(b)). "The Act provides exemptions to protect, for example, national security, personal privacy, privileged records, and law enforcement interests"[6]. Under Rule 83 (17 CFR 200.83), which defines the Confidential treatment of files, requires a party to "mark each page with "Confidential Treatment Requested by [name]" and an identifying number and code, such as Bates-stamped number"[7].

Ensuring that each page produced in legal proceedings contain the correct Bates numbers and Confidentiality stamping in response to investigations or litigation is critical. Bates numbers are relied upon in written submissions to the courts, used to refer to specific pages within the production at both hearings and depositions. Confidentiality designations are relied upon to ensure the appropriate protections are applied to each document and page produced in order to appropriately limit the disclosure of potentially sensitive information. Although document production software exists that can apply the Bates stamps and other Confidentiality or FOIA stamping in an automated process, it is sometimes encounters errors. A disruption in a network connection or running multiple competing processes concurrently can cause stamping software to inadvertently miss applying the proper numbering or designation to an image. Upon completion of automatic stamping, it is best practice to apply a quality control process to ensure that images have been properly stamped. This quality control process is manual, but in a large document production with hundreds of thousands of pages, looking at every page isn't feasible., Instead, a person typically reviews a sample of images to confirm the stamping was properly applied.

## II. DEEP LEARNING AND OCR

Optical character recognition (OCR) is a method that recognizes texts from image format files that do not have selectable or searchable text such as scanned PDFs, or TIFFs. Traditional OCR uses patterns and correlation to differentiate words from other visual elements of the file. With the advances of deep neural networks in image analysis, deep learning techniques have been applied recently to OCR [12]. There are various deep learning models that have been used in OCR, including long short-term memory (LSTM) and attention-based models. For our Bates number recognition, we used Tesseract OCR that utilizes LSTM. (Figure 1)

Tesseract OCR is an open source tool that was originally created at Hewlett-Packard (HP) in the 1980's. Tesseract was open sourced later and further developed by Google [8]. In 2018, a LSTM neural network model was introduced to the Tesseract OCR engine [9]. Tesseract OCR is currently a popular free OCR tool used for text detection and recognition.

Tesseract OCR is implemented in C++. It provides both command line program and application programming interfaces (APIs). We use *tesserocr* [10], a Python wrapper of the API for extracting Bates numbers. The wrapper provides tools to enable Python parallel processing, which is essential for scalability with large number of images.

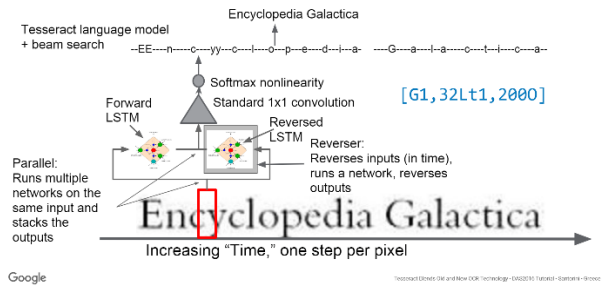

*Figure 1. Illustration of LSTM in Tesseract v4 [9]*

## III. THE APPROACH

A typical Bates stamp is shown in Figure 2, an image where the left bottom corner shows a Confidentiality stamp and the right bottom corner is the Bates number stamp. The extraction of the stamps consists of the following steps:

1. Crop the image to keep the bottom region of a predefined height as shown in Figure 3. This preprocessing is more efficient than feeding the whole image to the OCR engine. The specific height can be configured to avoid the stamps being cropped out.
2. Split the region into left and right corners with two thirds of the width as left corner and one third for the right corner. The right corner is for the Bates numbers, and the left corner is for Confidentiality stamps, which are typically more characters than a Bates number.
3. Feed the two corner regions to Tesseract OCR engine with API call. The API returns the text it. The

returned results are then used for either stamp detection or text extraction based on the use cases.
4. Occasionally, specific legal matters specify that documents have the stamps at the top instead or in addition to the bottom. In these instances, the same action can be performed on the top portion of the page to extract potential left and right corner stamps.
5. Apply post-processing to the OCR outputs to correct common errors in the output. For example, the Bates number prefix provided by the user can be used to check against the extracted text, and number of digits of the Bates number extracted can be counted. This information can be used in turn to fix incorrectly extracted text. Other examples include number 0 being misread as letter O, and a space being inserted in the Bates number, which can be fixed in the post process.

## IV. PARALLEL PROCESSING FOR SCALABILITY

In eDiscovery, it is common to have hundreds of thousands of documents in a case, potentially resulting in millions of pages to check. The Tesseract OCR can be parallelized using Python multiprocessing. To do that, a multiprocessing pool is created with the desired number of CPU cores and the input images are automatically pooled into each process, which invokes an OCR engine and uses it to extract the text in batches for the pooled images.

## V. ACCURACY

We applied the solution outlined above to the quality control of an eDiscovery production. The project contains over 186,000 document images, each has a Bates Number metadata field that accompanies the production that should match the Bates number, if stamped and recognized correctly. This attribute enabled us to check the accuracy rate of the Bates number extraction process by comparing the metadata with the extracted Bates number for each document. The findings are as follows:

With the raw output of the Tesseract OCR engine, the matching rate is at only 56%, However, upon inspecting the mismatch items, we found the errors fall into several patterns as follows:

1. There are new lines before or after a Bates number due to letters above or below the Bates number
2. A space is inserted into a Bates number
3. A number 0 is misread as letter O
4. Bates prefix has an additional letter O

With these patterns identified, the errors were easily fixed with a post processing function. The results after the postprocessing led to near 100% accuracy rate with only 9 mismatches left. Among the 9 mismatches, 7 images have no Bates number extracted at all, while the other two have incorrect extraction that do not follow a generic pattern to be fixed as part of post processing. All of the 9 documents do have correct Bates stamps upon manual inspection.

In our experiment data set, every document image contained a Bates number, so the extraction process provides validation of the Bates stamping. On the other hand, for the same data set, the extraction of Confidentiality stamps found that about 8 percent of the documents did not contain Confidentiality stamps. This provided the user a quicker way to find potential errors in the stamping rather than manually sampling the documents for the same purpose, with the potential that the sampling process didn't include any of these documents for review.

It should be pointed it out that this is just one case for the accuracy observation. While different image creation and stamping engines have different level of image quality and OCR output quality, the general post process concept should be applicable on a case by case basis to boost the accuracy rate.

## VI. ADDITIONAL CONSIDERATIONS AND FUTURE WORK

An important aspect of OCR pipeline to boost the accuracy rate is image preprocessing, including rescaling, noise removal, rotation/de-skewing, etc. [11]. Additionally, in the post processing step, more sophisticated analytic techniques such as statistics can also be used as needed.

For future work, we will address two other errors seen in legal productions. The first enhancement will be to check that the Bates numbers have been applied sequentially. When there is a number in the Bates sequence missing, it's considered a Bates gap and indicates an omission in the production. Most Bates stamping issues in practice are that the stamping engine somehow gets tripped up and applies the wrong number rather than not applying a stamp at all. This process can be used in conjunction with metadata included in a document production to validate the results.

The second future application is to identify items where there is more than one stamp, or sometimes referred to as a double stamp. This can occur when an image has been stamped and needs to be included in a new production or reproduced to a new party preserving the original stamp, or needs to resolve an error with the creation of the image causing the system to add a second Bates number to the image. Future work can be undertaken to identify these instances.

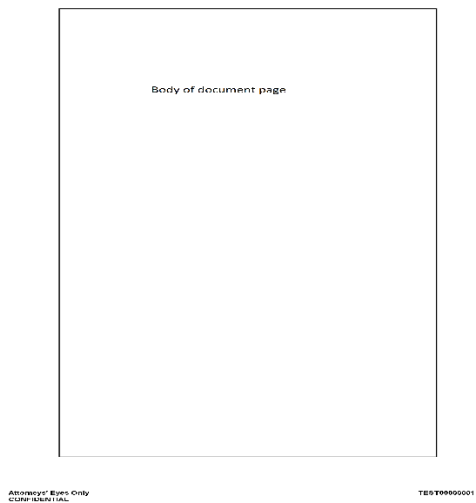

Figure 2. an illustration of document stamping, where the left bottom corner is Confidentiality and the right is the Bates number

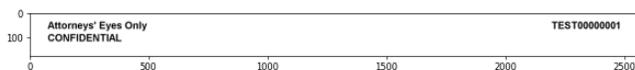

Figure 3. cropped region for OCR